\documentstyle[12pt]{article}
 \def\NPB#1#2#3{Nucl. Phys. B {\bf#1} (19#2) #3}
 \def\PLB#1#2#3{Phys. Lett. B {\bf#1} (19#2) #3}
 
 \def\PRD#1#2#3{Phys. Rev. D {\bf#1} (19#2) #3}
 \def\PRL#1#2#3{Phys. Rev. Lett. {\bf#1} (19#2) #3}

 \def\ARNP#1#2#3{Ann. Rev. Nucl. Part. Sci. {\bf#1} (19#2) #3}

 \def\beq{\begin{equation}}
 \def\eeq{\end{equation}}
 \newcommand{\newc}{\newcommand}
 \newc{\sm}{Standard Model }
 \newc{\smd}{Standard Model}
 \newc{\dac}{discrete anomaly cancellation }
 \newc{\mup}{``$\mu$'' problem }
 \newc{\eps}{\epsilon}
 \newc{\barr}{\begin{eqnarray}}
 \newc{\earr}{\end{eqnarray}}
 \def\gappeq{\mathrel{\rlap {\raise.5ex\hbox{$>$}}
 {\lower.5ex\hbox{$\sim$}}}}
 \def\lappeq{\mathrel{\rlap{\raise.5ex\hbox{$<$}}
 {\lower.5ex\hbox{$\sim$}}}}
\def\l{\label}
\def\r{\ref}

\begin{document}


\begin{flushright}
 CERN-TH-95-133 \\
 HD-HEP-95-8 \\
 IOA-95-323 \\
\end{flushright}

 \begin{center}
 {\bf HEAVY NEUTRINO THRESHOLD EFFECTS IN  LOW ENERGY
PHENOMENOLOGY} \\
 \vspace*{1.0cm}
{\bf G. K. Leontaris$^{1,2}$,
  S. Lola$^{3}$,} and
{\bf  G. G. Ross$^{4,5\star}$}
 \end{center}
 \vspace*{0.1cm}
 \begin{center}
 \begin{tabular}{l}
$^{1}${\small Centre de Physique Theorique, Ecole Polytechnique,
  F-91128 Palaiseau, France}\\
 $^{2}$
 {\small {\it on leave from}
Physics Department, Ioannina University, Ioannina, Greece}\\
 $^{3}${\small Institut f\"{u}r
 Theoretische Physik, Univerisit\"at Heidelberg,}\\
 {\small Philosophenweg 16, 69120 Heidelberg, Germany }\\
 $^{4}${\small Department of Physics,Theoretical Physics, University of
 Oxford,}\\
 {\small 1 Keble Road, Oxford OX1 3NP}\\
 $^{5}${\small CERN Theory Division, Geneva , Switzerland}\\
 \end{tabular}
 \end{center}

 \begin{center}
 {\bf ABSTRACT}
 \end{center}

\noindent
\small{
Right handed neutrinos with  mass of
${\cal O} \; (10^{12}-10^{13})$ GeV are  required to implement
the see-saw mechanism and generate
neutrino masses capable of playing a role in structure formation.
 Moreover models of fermion masses often relate the Yukawa couplings
involving these neutrinos to the up-quark  Yukawa couplings. Here we study
the effects of such couplings on the radiative corrections to quark masses.
We find that  $b-\tau$ equality at $M_{GUT}$ may still give the correct
$m_b/m_{\tau}$-- ratio at low energies, but only if there is large $\mu-\tau$
mixing  in the charged leptonic sector. We propose specific mass matrix
``textures'' dictated by a $U(1)$ family symmetry    whose structure
preserves  $m_b=m_{\tau}$ at $M_{GUT}$.  In these schemes, due to the large
$\nu_{\mu} - \nu_{\tau}$ mixing, it is possible to give a
simultaneous solution to the solar neutrino deficit
and the atmospheric neutrino problem.

 \noindent

 \noindent
 \rule[.1in]{14.5cm}{.002in}

 \noindent
 $^\star$ SERC Senior Fellow.

\thispagestyle{empty}

\setcounter{page}{0}
\vfill\eject

 \section{Introduction}

 Although the \sm is remarkably successful in describing strong and
electroweak  phenomena it still leaves many
 questions unanswered. In particular one would  like to understand the
 origin of fermion masses and mixing angles
 which in the \sm appear as arbitrary
 parameters. An obvious possibility is that some structure
 additional to that of the \sm
 is responsible for the
 pattern of masses and mixings
 that we see at low energies.
Support for such a stage of unification has been obtained for, although
unification \cite{unif,unif1}
 on its own does not agree with
 experiment, when combined with supersymmetry
 it leads to very successful predictions \cite{1a}
 for
 the  gauge couplings, the
 pattern and magnitude of spontaneous symmetry
 breaking at the elecroweak scale \cite{radi}
 and the bottom -- tau ($ b$ --  $\tau$)  unification \cite{1a,1}.
 A further indication that
 additional symmetries beyond the \sm
 exist, has been the observation that
 the fermion mixing angles and masses have values
 consistent with the appearance of
 ``texture'' zeros in the mass
 matrix \cite{3,3b,7,7b,IR,DLLRS,brn}.
 Such zeros may indicate the presence of an additional broken symmetry.
 When
unbroken only
 the third generation is
 massive and all mixing angles are zero.
 However,
 symmetry breaking terms gradually fill in the mass matrix
 and generate a hierarchy of mass scales and
 mixing angles.

In this paper we will consider the implications for
 the neutrino sector extending the analysis presented
 in \cite{DLLRS}. We shall consider models with
 right-handed neutrinos in which both Dirac and Majorana masses
 for the neutrinos are present, and
 the ``see--saw'' mechanism
 automatically explains the
 lightness of neutrinos relative to the
 charged fermions.

 Neutrino data from various experiments can be explained
 if there is mixing between the various types of neutrinos.
 The solar neutrino puzzle can be resolved through
 matter enhanced oscillations (preferably between $\nu_e$
 and $\nu_{\mu}$ states) with a mixing angle somewhat smaller
 than the $1/3$ of the corresponding  Cabbibo
 angle of the quark sector, $V_{CKM}^{12}$. For this explanation to work,
the
squared
 mass difference of the two types of neutrinos involved in this
 phenomenon should lie in a very narrow region.
  The specific ranges for the angle and mass squared given by
 the latest experimental data are:

 For matter enhanced oscillations in the sun:
 \beq
 \sin2\theta_{ex} = (0.39-1.6)\times 10^{-2} ,
 \quad \Delta m^2= (0.6- 1.2)\times 10^{-5}
 eV^2, \label{eq:sol1}
 \eeq
 and for vacuum oscillations:
 \beq
 \sin2\theta_{ex} \ge 0.75 ,
 \quad \Delta m^2= (0.5-1.1)\times 10^{-5} eV^2.
 \label{eq:sol2}
 \eeq
 If we wish to avoid fine tuning problems, it seems necessary to
 assume that such small differences in neutrino masses can
 be obtained only if the $\nu_{e,\mu}$  neutrino
 masses themselves are also of the same order.  Finally,
if neutrinos play a role
in structure formation, providing a hot dark matter component,
 then the heavier neutrino(s) should have  mass in the
 range  $\sim (1-6)$ eV, the precise value depending on the number
of neutrinos
that have masses of this order of magnitude.

What symmetry could explain this pattern of masses?
If gauge symmetries have something
to do with
 the hierarchical  fermion mass
 spectrum a similar hierarchy may be expected to hold for
 the unknown neutrino masses too.
 In a previous study it was found that the observed
 hierarchical mass spectrum of the charged fermions
 (quarks and leptons) follows naturally if we extend the
 gauge group of the minimal supersymmetric standard model
 by a $U_{FD}(1)$ family type symmetry.
 The extension of this model to include the right
 handed  neutrino in the theory resulted in a similar
 structure of the neutrino sector as well
 leading to the following general structures \cite{DLLRS}:

  {\it 1.)} The solar neutrino puzzle can be explained via
     $\nu_e \rightarrow \nu_{\mu}$ oscillations.
     The hierarchical mass spectrum leads to the conclusion
     that $m_{\nu_{\mu}}\approx \sqrt{\delta m_{\mu,e}^2}$
     while $m_{\nu_{e}}\ll m_{\nu_{\mu}}$. This also fixes
     the right handed neutrino scale $M_N$ through the effective
     Majorana mass matrix resulting from the usual ``see -- saw''
     mechanism
 \beq
 m^{eff}_{\nu}=-\frac{1}{4}m_{\nu_D}M_N^{-1}
                                  m_{\nu_D}^T\label{eq:meff}
 \eeq

  {\it 2.)}  The simultaneous solution of the solar neutrino problem
 and the interpretation of the $\nu_{\tau}$ mass  as a hot
 dark matter component through the effective light Majorana
 neutrino mass matrix, requires a right handed neutrino scale
 of the order of $M_{\nu_R}\sim 10^{12}-10^{13}$ GeV.

  {\it 3.)} There is no natural solution of the atmospheric neutrino
problem unless a considerable fine-tuning of the coefficients
in the neutrino mass textures occurs. This follows because the U(1) symmetry
that was
used to derive the above textures together with simple spontaneous breaking
gives only a hierarchical mass
spectrum and small mixing angles for all fermion mass matrices.

One additional implication of the structure emerging
from the $U(1)_{FD}$ symmetry is that the right handed neutrinos have Yukawa
couplings of the same order as the up quarks. This in turn affects the
radiative corrections in the model and in particular the expectations for
gauge unification and for the $m_b/m_{\tau}$ ratio. The implications of such
large couplings has already been explored in refs.\cite{VB}. Here we develop
these analyses in two respects. Firstly we present a semi-analytic analysis
of the radiative corrections that allows us to analyse the possibilities for
maintaining $b - \tau$ equality at the GUT scale even in the presence of these
radiative corrections through large $\mu - \tau $ mixing giving a mechanism
to evade conclusion 3.) above\footnote{For another mechanism evading this
result through the introduction of spontaneous $U(1)$ breaking via several
Higgs scalars see ref.\cite{elen}.}.
 This leads us to consider schemes based on
the $U(1)$ family symmetry which naturally generate such mixing.
 The implications for the neutrino mass spectrum are then explored.
In addition the semi-analytic approach is supported by a full numerical
calculation. In section 2 we give a semi--analytical
approach to the renormalisation group
equations, in the presence of
right handed neutrinos.
These equations are used in section
3 in order to get some direct intuition on the
effects of the heavy neutrinos. The explicit
form of the solutions makes it easy to
see how $b-\tau$ equality at a GUT scale may be
made consistent with the parameter spectrum at low
energies, by sufficient
$\mu-\tau$ mixing
in the charged leptonic sector
and for a relatively heavy strange quark.
This is discussed in section 4.
In section 5 we give the resulting predictions
for the heavy and light Majorana neutrino
mass matrices and eigenvalues, with a mixing
which is of the correct order of magnitude
in order to explain the atmospheric
neutrino problem.
In section 6 we present a numerical approach
to the renormalisation group equations, which
depicts the observations of section 2.
Finally, in section 7 we give a summary of our results.

 \section{RGE with RH-neutrinos: a semi -- analytic approach}

 {}From the above it is clear that the interpretation of many
 important experimental facts is based on the existence of
 the right -- handed partners $\nu_{R_i}$ of the three left -- handed
 neutrinos, where the scale of mass of these particles is at least
three orders of
magnitude smaller
 than the gauge unification scale, $M_U$.
  Thus the running from the Unification scale, $M_U\sim
 10^{16}$ GeV, down to the scale of $M_{\nu_R}$, must include
 radiative corrections  from  $\nu_R$ neutrinos. After that
 scale, $\nu_R$'s decouple from the spectrum, and an effective
 see -- saw mechanism is operative, c.f. eq( \ref{eq:meff}).

 In the presence of the right handed neutrino, the renormalization
 group  equations for the Yukawa couplings at the one--loop level
 are
 \barr
 16\pi^2 \frac{d}{dt} h_U&=
 & \left(
 3 h_U h_U^\dagger +h_D h_D^\dagger
 +I\cdot
 Tr[h_N h_N^\dagger]
 +  I\cdot Tr[3 h_Uh_U^\dagger ]
             - I\cdot G_U \right)  h_U, \label{eq:rge1}
\\
 16\pi^2 \frac{d}{dt} h_D&=
 & \left(
 3 h_D h_D^\dagger +h_U h_U^\dagger
  +I\cdot Tr [3 h_Dh_D^\dagger] + I\cdot Tr[h_E
 h_E^\dagger ] - I\cdot G_D \right) h_D, \label{eq:rge3}
 \\
 16\pi^2 \frac{d}{dt} h_E&=&\left(
  3 h_E h_E^\dagger+h_N h_N^\dagger +
  I\cdot Tr [ h_Eh_E^\dagger] +
 I\cdot  Tr[3 h_D  h_D^\dagger ]
  -  I\cdot G_E \right) h_E,
 \label{eq:rge4} \\
 16\pi^2 \frac{d}{dt} h_N&=& \left(
 h_E h_E^\dagger + 3h_N
 h_N^\dagger
 + I\cdot Tr [3 h_U h_U^\dagger  ]
  + I\cdot Tr[ h_N h_N^\dagger] - I\cdot G_N \right) h_N.
 \label{eq:rge2}
 \earr
 where $h_\alpha$, $\alpha=U,D,E,N$, represent the
 $3 \otimes 3$ Yukawa matrices for the up and down quarks, charged
lepton and Dirac neutrinos, while $I$ is the $3 \otimes 3$ identity
matrix. Finally, $G_{\alpha}= \sum_{i=1}^3c_{\alpha}^ig_i(t)^2$ are
functions which depend on the  gauge couplings with the
coefficients $c_{\alpha}^i$'s given by \cite{DLT,VB}.
\barr
\{c_U^i \}_{i=1,2,3} &=& \left\{ \frac{13}{15},3,\frac{16}{3}
\right\}, \qquad \{c_D^i \}_{i=1,2,3} =
\left\{\frac{7}{15},3,\frac{16}{3} \right\}, \\  \{c_E^i
\}_{i=1,2,3} &=& \left\{ \frac{9}{5},3,0\right\}, \qquad \quad
\{c_N^i \}_{i=1,2,3} = \left\{ \frac{3}{5},3,0\right\}.
\earr

 Consider  initially the simple case where only the top and Dirac --
 type neutrino Yukawa couplings are large at the GUT scale
 (i.e. the case of small $tan\beta$ scenario).
 Let us start assuming that the top and neutrino
 Yukawa couplings are equal at the
 Unification scale, $h_t(M_U)=h_N(M_U)$, a relation which
 arises naturally not only in our case but in most of the Grand
 Unified Models which predict the existence of the right handed
 neutrino.
 As in the case of the charged fermions, we will consider only
hierarchical textures \cite{DLLRS} for the right handed neutrino
Majorana mass matrices, i.e.  $M_{\nu_1}\ll M_{\nu_2}\ll
M_{\nu_3}$.
 If we work in a diagonal basis  we can considerably simplify
  the above equations. Then, for the range $M_U$ to $M_N$,
 the renormalization group evolution of the Yukawa couplings of
 third generation, can be written as follows
 \barr
 16\pi^2 \frac{d}{dt} h_t&=
 & \left(
 6 h_t^2  + h_N^2
   - G_U\right)  h_t, \label{eq:rg1}
 \\
 16\pi^2 \frac{d}{dt} h_N&=& \left(
  4h_N^2  + 3 h_t^2
   - G_N \right) h_N,
 \label{eq:rg2}  \\
 16\pi^2 \frac{d}{dt} h_b&=
 & \left(h_t^2 - G_D \right) h_b, \label{eq:rg3}
 \\
 16\pi^2 \frac{d}{dt} h_{\tau}&=&\left( h_N^2
  - G_E \right) h_{\tau},
 \label{eq:rg4}
 \earr
 Below $M_N$, the right handed neutrino decouples from the
 massless spectrum and we are left with the standard spectrum
 of the MSSM.  For scales $Q\le M_N$ the gauge and Yukawa
 couplings evolve according to the standard renormalisation
 group equations. In addition, the effective Yukawa coupling
 of the light neutrino mass matrix (\ref{eq:meff}) evolves
 according to \cite{BP}
  \begin{eqnarray}
 16\pi^2 \frac{d}{dt}
 h_{\nu} &= h_{\nu} (I \cdot
 Tr[6 h_Uh_U^{\dagger}]-G_{\nu})+ h_{\nu}
 h_Eh_E^{\dagger}+ h_E^{\dagger}h_E
 h_{\nu}
 \end{eqnarray}
 with $G_{\nu}= 2 g_1^2 +6 g_2^2$.
 In order to gain an insight into the effects of new couplings
 associated with  the
$\nu_R$ in the renormalisation group running we
integrate the above equations in the region
$M_N\le Q\le M_U$. We denote the top and $\nu_R$ Yukawas by
$h_G$ at the  unification scale, while bottom and
 $\tau$ are denoted with $h_{b_0},{h_{\tau_0}}$
respectively. We get
  \begin{eqnarray}
 h_t(t)&=&\gamma_U(t)h_G\xi_t^6\xi_N\\
 h_N(t)&=&\gamma_N(t)h_G\xi_t^3\xi_N^4\\
 h_b(t)&=&\gamma_D(t)h_{b_0}\xi_t\\
 h_{\tau}(t)&=&\gamma_E(t)h_{\tau_0}\xi_N
 \end{eqnarray}
 where the functions $\gamma_\alpha(t)$ depend purely on
 gauge coupling constants and are given by
 \barr
 \gamma_\alpha(t)&=&  \exp({\frac{1}{16\pi^2}\int_{t_0}^t
  G_\alpha(t) \,dt})\\
 &=& \prod_{j=1}^3 \left( \frac{\alpha_{j,0}}{\alpha_j}
 \right)^{c_\alpha^j/2b_j}
 \\
 &=& \prod_{j=1}^3 \left(1- \frac{b_{j,0}\alpha_{j,0}(t-t_0)}
 {2\pi}\right)^{c_\alpha^j/2b_j},
 \earr
 The $\xi_i$'s ($i=t,N,b,\tau$) are given by the integrals
 \barr
 \xi_i&=& \exp({\frac{1}{16\pi^2}\int_{t_0}^t \lambda^2_{i}dt})
 \earr
 The values of the parameters $\xi_i$ can be determined
 at any scale by numerically solving  the renormalization group
 equations.
 As a general remark, we note that the initial condition for
 $\xi_i$ is  $\xi_i(t_U)=1$, while at any lower scale $Q<M_U$,
 $\xi_i(Q)<1$.

 \section{Heavy neutrino effects : an insight}

 We start by investigating the $b-\tau$ Yukawa coupling
 solutions.
Thus, in the case of small $tan\beta$, we can relate their
 solutions  at the scale $M_N$
 in terms of the initial values, by the following equation
 \begin{equation}
 h_{b}(t_N)=\rho
 \xi_t\frac{\gamma_D}{\gamma_E}h_{\tau}(t_N) \label{rho}
 \end{equation}
 with $\rho=\frac{h_{b_0}}{h_{\tau_0}\xi_N}$.
 In the case of $b-\tau$ unification at $M_U$, we have
  $h_{\tau_0} =h_{b_0}$, while in the absence
 of the right -- handed neutrino $\xi_N \equiv 1$, thus
 $\rho =1 $ and the $m_b$ mass has the phenomenologically reasonable
prediction at low
 energies given by the approximate one-loop formula
 \begin{equation}
 m_{b} = \eta_b
 \xi_t \frac{\gamma_D}{\gamma_E}  m_{\tau}
 \end{equation}
where $\eta_b$ is the renormalization group coefficient in
the $m_t$--$m_b$ range.
 In the presence of $\nu_R$ however, if $h_{\tau_0}
 =h_{b_0}$ at the GUT scale, the parameter $\rho$
 is no longer equal to unity since $\xi_N<1$. In fact the
 parameter $\xi_N$ becomes smaller for lower $M_N$ scales.
 Therefore in order to restore the correct $m_b/m_{\tau}$ prediction at low
 energies we need $\rho =1$ corresponding to
 \begin{equation}
 h_{b_0}=h_{\tau_0}\xi_N
 \end{equation}
 Hence it is obvious that we need a $\tau -$Yukawa coupling
 $h_{\tau_0}$,
 larger than $h_{b_0}$ at $M_U$ to compensate
 for the factor $\xi_N$ arising from the presence of $\nu_R$.

 It is interesting that $\xi_N$ can be given in this case
 in terms of the values of gauge and Yukawa
 ratios at $M_N$ only, irrespective of the initial conditions
 \begin{eqnarray}
 \xi_N&=&\frac{h_{b_0}}{h_{\tau_0}}\nonumber \\
 &=&
  \frac{h_{b}\gamma_E}{h_{\tau}\gamma_D}
 \bigl(\frac{h_t\gamma_N}{h_N\gamma_U}\bigr)^{-1/3}
 \end{eqnarray}

 On the other hand, the top mass at the scale $t_N=ln(M_N)$
 can also be expressed formally as follows
 \begin{equation}
 m_{t}(t_N) =h_G\gamma_U(t_N)
 \xi_t^6(t_N)\xi_N(t_N)\frac{\upsilon}{\sqrt{2}}sin\beta
 \end{equation}
 where $\upsilon =246GeV$ and $tan\beta$ is the ratio of the
 two Higgs vev's. Again, in the absence of $\nu_R$ this
 reduces to the well known one loop approximate formula
 which coincides with the above for $\xi_N=1$.
 In the present case however, this prediction
  corresponds effectively to a smaller initial value
 of the top Yukawa coupling of the order
 \begin{equation}
 h_G^{\prime}=h_G \xi_N(t_N)
 \end{equation}
{}For $h_G> 1$, however, due to the infrared fixed point
property of the top -- Yukawa solution \cite{PR},
 the $m_t$ -- prediction is
not going to alter significantly. For the same $tan\beta$,
one will get almost the
previous top mass prediction, reduced at most by 2\%.
In contrast, in the small $tan\beta$
scenario where $h_b\ll 1$, one naturally expects that bottom
mass will be affected by the presence of $\nu_R$.
 For $M_N\approx 10^{13}GeV$ for example and
 $h_G \ge 1$, we can estimate that
 $\xi(t_N)\approx 0.89$ thus, there is a corresponding
 $\sim 10\%$ deviation of the $\tau - b$ equality at the
 GUT scale.

{}Furthermore,
we have seen that in order to recover the correct $m_b/m_{\tau}$
relation at low energies, it is necessary that $h_{b,0}/h_
{\tau ,0} < 1$ as long as $M_N < M_U$.
This can be done in two ways:
Either we can keep the same value of the $b$ -- Yukawa and increase
the $\tau$-Yukawa by a factor $\xi_N^{-1}$, or decrease the
bottom coupling by a factor $\xi_N$. In the first case, the angle
$\beta$ remains the same and the top mass unaffected.
In the second case, in order to retain the same absolute value
of the initial parameters for the $b,\tau$ masses we need to
increase $cos\beta$. This results to a corresponding decrease
of the top mass prediction.

We will present a detailed numerical analysis of the above in
section 5, where two loop effects from the gauge
couplings are taken into account. In the next section we first
propose a scheme  in which the bottom-tau unification
is restored.  We will show that this may result in
a solution of the solar neutrino deficit
and the atmospheric neutrino problem.

\section{Restoration of bottom -- tau unification}

Given the results of section 3, it is natural to ask if
 Grand Unified models which predict the $b - \tau$
equality at the Unification scale,  exclude the experimentally
required and cosmologically interesting
region for the neutrino masses.
To answer this question, we should first recall that
the $b-\tau$ -- equality at the GUT scale refers to the
$(33)$ entries of the corresponding charged lepton and
down quark mass matrices. The detailed structure of the
mass matrices is not predicted, at least by the Grand Unified
Group itself, unless additional structure is imposed.
It is possible then to assume $(m_E^0)_{33}=
(m_D^0)_{33}$  and a specific structure of the corresponding mass
matrices such that after the diagonalisation at   the
GUT scale, the $(m^{\delta}_E)_{33}$ and $(m^{\delta}_D)_{33}$
entries are no-longer equal.

To illustrate this point, let us present
 here a simple $2\times 2 $ example.
 Assume a diagonal form of $m_D^0$ at the GUT scale ,
 $m_D^0 = diagonal (c m_0,m_0)$, while the corresponding
 entries of charged lepton mass matrix have the form
 \beq
 m_{E}^0 =
 \left (
 \begin{array}{cc}
 d & \epsilon \\
 \epsilon &  1
 \end{array}
 \right) m_0
 \eeq
 These forms of $m_D^0,m_E^0$ ensure that at the GUT scale
 $(m_D^0)_{33}= (m_E^0)_{33}$. However, at the low energies
 one should diagonalize the renormalised Yukawa matrices
 to obtain the correct eigenmasses. Equivalently, one can
 diagonalise the quark and charged lepton Yukawa matrices
 at the GUT scale and evolve separately the eigenstates and
 the mixing angles. Since $m_D^0$ has been chosen diagonal
 there is no need of diagonalization and the mass eigenstates
 which are to be identified with the $s,b$ -- quark masses at
 low energies are given by
 \barr
 m_s=c \gamma_D m_0 ,&
 m_{b} =  \gamma_D  m_0 \xi_t
 \earr
 with $m_0 = h_{b_0} \frac{\upsilon}{\sqrt{2}}cos\beta$.
 To find the charged lepton mass eigenstates we need first to
 diagonalise $m_E^0$ at $M_{GUT}$. We can obtain the following
 relations between the entries $\epsilon ,d$ of $m_E^0$ and the mass
 eigenstates $m_{\mu}^0,m_{\tau}^0$ at the GUT scale.
 \barr
 d=(\frac{m_{\tau}^0 - m_{\mu}^0}{m_0}-1)\\
 \epsilon^2
  = (\frac{m_{\mu}^0}{m_0}+1) (\frac{m_{\tau}^0}{m_0}-1)
 \earr
 In the presence of right handed neutrino, the evolution of
 the above $\tau -$ eigenstate down to low energies is that
 described  by (\ref{eq:rg4}) with
 $m_{\tau_0}=h_{\tau_0}
 \frac{\upsilon}{\sqrt{2}}cos\beta$.  By simple comparison of
 the obtained formulae, we conclude that, to obtain the
 correct $m_{\tau}/m_b$ ratio at $m_W$ while preserving the
 $b - \tau$  unification at $m_{GUT}$, the $m_E^0$ entries
 should satisfy the following relations
 \barr
 \epsilon = \sqrt{\frac{1}{\xi_N}-1} ,&
 d \approx (\frac{1}{\xi_N}-1) = \epsilon^2
\l{eq:de}
 \earr
 The above result deserves some discussion.
 Firstly we see that it is possible to preserve $b - \tau$
 unification by assuming $2-3$ generation mixing in the lepton
 sector, even  if the effects of the $\nu_R$ states are included.
 Secondly, this mixing is related to a very simple parameter
 which depends only on the scale $M_N$ and the initial
 $h_N$ condition.
 The range of the coefficient $c$ in the diagonal form of the
$m_D^0$ -- matrix, can also be estimated
using the experimental values of the quark masses $m_s,m_b$.
An interesting observation is that the usual $GUT$ -- relation
for the $(22)$ -- matrix elements of the charged lepton and down quark
mass matrices, i.e., $(m_E)_{22}=-3 (m_D)_{22}$, which in our case
is satisfied for $c = -d/3$, implies here a relatively heavy strange
quark mass $m_s\sim 200$MeV.
Smaller $m_s$ values are obtained if $-3c/d <1$.
\footnote{An alternative  mechanism  which  restores  the correct
$m_b/m_{\tau}$ ratio in the presence of  $\nu_R$
 was proposed in \cite{dimp}.}

We turn now to a consideration whether the hierarchical structure of the
lepton
mass matrix corresponding to eq(\r{eq:de}) can be obtained
 by a simple $U(1)$
symmetry \cite{IR}.

In \cite{IR} a viable fermion mass matrix
 structure was constructed following from
a spontaneously broken $U(1)$ gauge symmetry. In this the form of the down
mass matrix is

\beq
\frac{M_d}{m_b} \approx
\left(
\begin{array}{cc}
\bar{\epsilon}^2 & \bar{\epsilon} \\
\bar{\epsilon} & 1
\end{array}
\right)\l{eq:md}
\eeq
(We have concentrated for simplicity  in the case of
the two heavier generations which are relevant here.)  Note that we have
suppressed all Yukawa couplings in writing this mass
 matrix -- only the order of
the matrix elements allowed by the broken symmetry are given.
These Yukawa
couplings are assumed to be of order 1 and the object
of the exercise is to
demonstrate that the hierarchical structure of the fermion masses
may come only
from symmetry considerations. Eq(\r{eq:md}) is diagonalised
 by the orthogonal
matrix
\beq
V \approx
\left (\begin{array}{cc}
1 & \bar{\epsilon} \\
-\bar{\epsilon} & 1
\end{array}
\right)
\eeq
where
$\bar{\epsilon} = 0.23$, in order to fit the
down quark masses and mixing angles.

The structure of the lepton mass matrix following from the $U(1)$ symmetry
(again for the heavier generations) is
\beq
\frac{M_l}{m_b} \approx
\left (
\begin{array}{cc}
\bar{\epsilon}^{2\mid \beta \mid} & \bar{\epsilon}^{\mid\beta\mid} \\
\bar{\epsilon}^{\mid \beta\mid } & 1
\end{array}
\right)
\l{eq:bt}
\eeq
where $\beta \equiv 1-b = \frac{a_2-\alpha_1}{\alpha_2-\alpha_1}$.
and in \cite{IR}, \cite{DLLRS}
the cases $\beta =1/2$ and $\beta=1$ had been considered.
Both possibilities are in very good agreement with the measured
masses and mixing angles.
 The important fact here is that
 $\beta $ can be determined by the requirement that
 $b - \tau$ mass ratio be correctly given when heavy neutrinos, which become
massive at an intermediate scale $M_{N}< M_U$,
are present. Allowing for the unknown coefficients
of $O(1)$ we see (cf.  eqs(\r{eq:bt}) and (\r{eq:de}))
 that both $\beta=1/2$ and $\beta=1$ are in reasonable
 agreement with the above expectation\footnote{Here we assume
the field spontaneously breaking $U(1)$ carries half -- integral $U(1)$ charge
so we do not have the $Z_2$ symmetry of \cite{IR}.}.
 Now the diagonalising matrix
is given by
\beq
V\approx
 \left(
\begin{array}{ccc}
\sqrt{1-\bar{\epsilon}^2} &
\bar{\epsilon}  \\
-\bar{\epsilon} & \sqrt{1-\bar{\epsilon}^2}
\end{array}
\right)
\eeq
Obviously, there is a large mixing in the $2-3$ lepton sector of
the obtained solution which may lead to interesting effects in
the rare processes like $\tau \rightarrow \mu \gamma$ and
neutrino oscillations. In the simplest realisation of this scheme
we expect $h_b\approx h_{t}$ because in the limit $\epsilon$, $\bar \epsilon
\rightarrow 0$ the $U(1)$ quantum numbers of the light Higgs $H_{1,2}$ allow
them to couple to the third generation and a $SU(2)_l\otimes SU(2)_R$
symmetry of the couplings ensures equal Yukawa coupling. Thus this model
applies only to the large $\tan \beta$ regime. However if there is an
additional heavy state, $H_i,\;\bar H_i,\; i=1 $ or  2, with the same $U(1)$
quantum number then mixing effects can generate different $h_b$ and $h_{t}$
couplings allowing for any value of $\tan \beta$.

\section{The Effective Light Majorana Mass Matrix}

We have seen in section 3, that we can obtain with a $U(1)$ family
symmetry a
charged lepton
mass matrix with the required large mixing in the two heavier
generations by choosing the one free parameter, $b$.
The choice $b=1/2$  gives a very good agreement
with the
charged lepton masses and the bottom-tau relation in the
presence of $\nu_R $ with mass $M_N\approx 10^{13}GeV$.
Our next step is to determine the Dirac and heavy Majorana
mass matrices. The general form of the Dirac neutrino mass
 matrix
 for arbitrary $\alpha ,\beta \equiv 1-b$ is given by \cite{DLLRS}
\beq
M^D_{\nu} =
\left (
\begin{array}{ccc}
\eps^{2\mid 3\alpha + \beta\mid } & \eps^{3\mid\alpha\mid}
&\eps^{\mid 3\alpha + \beta\mid } \\
\eps^{3\mid \alpha \mid } & \eps^{2\mid \beta \mid} &
\eps^{\mid \beta \mid}\\
\eps^{\mid 3\alpha + \beta\mid } & \eps^{\mid \beta \mid} & 1
\end{array}
\right)
\eeq

The Majorana masses for the right -- handed neutrinos
are generated by
terms of the form $\nu_{R} \nu_{R} \Sigma$, where
$\Sigma$ is a singlet scalar field
--invariant under the $SU(3)\otimes SU(2)_L\otimes U(1)_Y$ gauge group--
with charge $Q_{\Sigma}$ under
the $U(1)_{FD}$ family symmetry.
For the various choices of
$Q_{\Sigma}$ we may then form the possible
``textures''
for the heavy  Majorana mass
matrix.
For example, when $Q_{\Sigma} = -2 a_{1}$, that is when
the field $\Sigma$ has the same charge with the
Higgs fields, only the $(3,3)$ entry
of the mass matrix appears for an
exact $U(1)$ symmetry
\footnote{The full anomaly free Abelian group $U(1)$
involves an additional family independent
component $U(1)_{FD}$.}
and is of order
unity.

However, at a later stage  the Abelian symmetry is
broken
by standard model singlet fields
$\theta$ and $\bar{\theta}$ with
$U(1)$ charge $\pm 1/2$, which acquire vacuum
expectation values along a D -- flat direction.
At this stage, invariant combinations involving the
$\theta$, $\bar{\theta}$ fields are generated, filling
the rest of the entries in the mass matrices in
terms of an expansion parameter
\cite{IR}.
Depending on which elements of
the Majorana mass matrix we require
to appear before the $U(1)$ symmetry breaking,
we can make six different
choices of the charge $Q_{\Sigma}$ which result to the
$M_{\nu_R}^{maj.}$ -- ``textures''  shown in Table 1.

\begin{table}
\centering
\begin{tabular}{|c|c|} \hline
$\left (
\begin{array}{ccc}
\bar\eps^{2\mid 3\alpha + \beta\mid } & \bar\eps^{3\mid\alpha\mid}
&\bar\eps^{\mid 3\alpha + \beta\mid } \\
\bar\eps^{\mid 3\alpha \mid } & \bar\eps^{2\mid \beta \mid} &
\bar\eps^{\mid \beta \mid}\\
\bar\eps^{\mid 3\alpha + \beta\mid } & \bar\eps^{\mid \beta \mid} & 1
\end{array}
\right)$
 &
$\left (
\begin{array}{ccc}
\bar\eps^{3\mid 2\alpha + \beta\mid } & \bar\eps^{\mid 3\alpha + \beta\mid}
&\bar\eps^{\mid 3\alpha + 2\beta\mid } \\
\bar\eps^{\mid 3\alpha + \beta\mid } & \bar\eps^{\mid \beta \mid} & 1\\
\bar\eps^{\mid 3\alpha + 2\beta\mid }  & 1 &\bar\eps^{\mid \beta \mid}
\end{array}
\right)$
\\ \hline
$\left (
\begin{array}{ccc}
\bar\eps^{2\mid 3\alpha + 2\beta\mid } &
 \bar\eps^{\mid 3\alpha + 2\beta\mid}
&\bar\eps^{3\mid \alpha + \beta\mid } \\
\bar\eps^{\mid 3\alpha + 2\beta\mid } &  1&\bar\eps^{\mid \beta \mid} \\
\bar\eps^{3\mid \alpha + \beta\mid }  &\bar\eps^{\mid \beta \mid}
&\bar\eps^{2\mid \beta \mid}
\end{array}
\right)$
 &
$\left (
\begin{array}{ccc}
\bar\eps^{\mid 3\alpha + \beta\mid } &
 \bar\eps^{\mid \beta\mid} &1 \\
\bar\eps^{ \mid\beta\mid } & \bar\eps^{3\mid \alpha + \beta\mid }&
\bar\eps^{3\mid \alpha + 2\beta\mid } \\
 1 &\bar\eps^{3\mid \alpha + 2\beta\mid }
 &\bar\eps^{\mid 3\alpha + \beta \mid}
\end{array}
\right)$
 \\ \hline
$\left (
\begin{array}{ccc}
1&\bar\eps^{\mid 3\alpha + 2\beta\mid } &
 \bar\eps^{\mid 3\alpha + \beta\mid}  \\
\bar\eps^{\mid 3\alpha +  2\beta\mid } &
 \bar\eps^{2\mid 3\alpha + 2\beta\mid }&
\bar\eps^{3\mid 2\alpha + \beta\mid } \\
\bar\eps^{\mid 3\alpha + \beta\mid} &
\bar\eps^{3\mid 2\alpha + \beta\mid }&
\bar\eps^{2\mid 3\alpha + \beta \mid}
\end{array}
\right)$
&
$\left(
\begin{array}{ccc}
\bar\eps^{\mid 3\alpha + 2\beta\mid }
& 1
 & \bar\eps^{\mid\beta\mid } \\
1  &
 \bar\eps^{\mid 3\alpha + 2\beta\mid }
 &\bar\eps^{\mid 3\alpha + \beta\mid } \\
\bar\eps^{\mid\beta\mid }  &
\bar\eps^{\mid 3\alpha + \beta\mid} &
\bar\eps^{\mid 3\alpha\mid}
\end{array}
\right)$
\\ \hline
\end{tabular}
\small{\caption{
General forms of heavy Majorana mass matrix textures.
The specific textures of the
text arise for $\alpha =1, \beta = 1/2$.}}
\label{table:maj}

\vspace{0.3 cm}

\centering
\begin{tabular}{|c|c|} \hline
$\left (
\begin{array}{ccc}
e^{10} &  &  \\
 & e^2 &  \\
 &  & 1
\end{array}
\right)$
 &
$\left (
\begin{array}{ccc}
e^{15} &  &  \\
 & -1+e &  \\
 &  & 1+e
\end{array}
\right)$
\\ \hline
$\left (
\begin{array}{ccc}
e^{16} &  &  \\
 & e^2 &  \\
 &  & 1
\end{array}
\right)$
 &
$\left (
\begin{array}{ccc}
e^9 &  &  \\
 & -1-e^2 &  \\
 &  & 1+e^2
\end{array}
\right)$
 \\ \hline
$\left (
\begin{array}{ccc}
e^{16} &  &  \\
 & e^{14} &  \\
 &  & 1
\end{array}
\right)$
&
$\left(
\begin{array}{ccc}
e^6 &  &  \\
 & -1-e^2 &  \\
 &  & 1+e^2
\end{array}
\right)$
\\ \hline
\end{tabular}
\caption{
Eigenvalues of Heavy Majorana mass matrix textures,
for $\alpha = 1$ and $\beta = 1/2$}
\label{table:majei}

\centering
\begin{tabular}{|c|c|} \hline
$\left (
\begin{array}{ccc}
e^{26} &  &  \\
 & e^{10} &  \\
 &  & 1
\end{array}
\right)$
 &
$\left (
\begin{array}{ccc}
e^{25} &  &  \\
 & e^9 &   \\
 &  & 1/e
\end{array}
\right)$
\\ \hline
$\left (
\begin{array}{ccc}
e^{24} &  &  \\
 & e^8 &  \\
 &  & 1/e^2
\end{array}
\right)$
 &
$\left (
\begin{array}{ccc}
e^{33} &  &  \\
 & e^{13} &  \\
 &  & 1/e^7
\end{array}
\right)$
 \\ \hline
$\left (
\begin{array}{ccc}
e^{40} &  &  \\
 & 1/e^8 &  \\
 &  & 1/e^{14}
\end{array}
\right)$
&
$\left(
\begin{array}{ccc}
e^{32} &  &  \\
 & e^{6} &  \\
 &  & 1/e^6
\end{array}
\right)$
\\ \hline
\end{tabular}
\caption{
Eigenvalues of light Majorana mass matrix textures,
for $\alpha = 1$ and $\beta = 1/2$}
\label{table:majeil}
\end{table}

For $\alpha =1, \, \beta = 1/2$, we can obtain the specific forms
for  Dirac and Majorana textures compatible with the correct
fermion mass predictions in the presence of the intermediate
neutrino scale.
In Table 2 we present the eigenvalues of
the heavy Majorana mass matrix for this choice
of $\alpha$ and $\beta$.

The analysis of the resulting $m_{\nu}^{eff}$
follows the same steps as in ref.\cite{DLLRS}.
In the matrices  we use
\begin{equation}
e = \bar{\epsilon}^{1/2}, \;
\bar{\epsilon} = 0.23
\end{equation}
The eigenvalues of $m_{eff}$ are given
in Table 3.
The order of the
matrices in Tables 2 and 3 corresponds to
the one of Table 1.

Note the interesting feature that
the large mixing in the (2,3) entries of the
charged leptons
  which we introduced
in order to restore the $b - \tau$ unification leads to a similar
large mixing in the
neutrino sector\footnote{The mixing in the (1,2) is of the ${\cal O}
((\frac
{m_e}{m_{\mu}})^{1/2})$  and negligible in  (1,3).}.
This mixing is of the correct order of magnitude
for a possible solution to the atmospheric
neutrino problem.
Indeed, the atmospheric
neutrino problem may be explained in
the case that large mixing and small
mass splitting involving the
muon neutrino exists \cite{atmo}.
Taking into account the
bounds from accelerator
and reactor disappearance
experiments one finds that
for $\nu_{e}-\nu_{\mu}$ or
$\nu_{\tau}-\nu_{\mu}$ oscillations
\begin{eqnarray}
\delta m^2_{\nu_{\alpha}\nu_{\mu}}&\geq&
10^{-2} \; {\rm eV}^{2}
\\
sin^22\theta_{\mu \alpha}&\geq& 0.51-0.6
\end{eqnarray}

  In \cite{DLLRS}, such a large
mixing was not present due to a residual discrete symmetry assumed
for the $b=1/2$ case. In that case, from the
resulting $m^{eff}_{\nu}$
mass matrix we had been able to fit the COBE
results
and solve the solar neutrino problem
(solution A).

In the case of the large mixing
discussed here we may also have a
similtaneous solution to the solar
and the atmospheric
neutrino problems (solution B).
However, the small mass splitting
which is required between
the neutrinos that mix in both the
solar and atmospheric neutrino oscillations,
make it impossible to
account for the COBE measurements at the same time.
This is a result of working in the minimal scheme
with only one $\Sigma$ field present, which
naturally leads to a large splitting
between the neutrino masses.
However, in the case that we add more
singlets $\Sigma$ in the theory,
it is possible to obtain
a heavy Majorana mass
matrix that leads to a solution
of all three problems similtaneously \cite{elen}.

Going back to the case of a single
$\Sigma$ field, whether we obtain
 the solution (A) or (B)
depends on the predicted mass splitting between
the two heavy neutrinos in
each of the six choices of the heavy Majorana mass matrix.
For a $\nu_{\tau} \approx 5$ eV and $x_{i} \equiv
 e^6$, $e^8$
and $e^{10}$, for
$i=1,2,3$,
 we obtain a muon neutrino mass
$m_{\nu_{\mu}} = m_{\nu_{\tau}} x_{i} = 0.06, 0.014$ and $0.003$
eV respectively. This indicates that
our solutions with a total splitting $e^{10}$ naturally
lead to a solution of the COBE measurements and
the solar neutrino problem.
On the other hand, for
$m_{\nu_{\tau}} \approx 0.1$ eV and  $x_{1} = e^6$,
$m_{\nu_{\mu}} = m_{\nu_{\tau}} x_{1} = 0.0012$ eV, which
may be marginally consistent with a solution to the
atmospheric and solar neutrino problems (remember
that coefficients of order unity have not
yet been defined in the solutions).
Since there are alternative schemes which lead to
an explanation of the COBE measurements, other
than hot and cold dark matter
\footnote{ For example, we have found that
domain walls may give structure at medium and
large scales if, either they are unstable, or the
minima of the potentials of the relevant scalar
field appear with different probabilities \cite{walls}.}
we believe that the scheme (B) should be
considered on equivalent grounds with the scheme (A).


 \section{Numerical analysis}

 In this section, we present the results of a numerical analysis of
the effects of the heavy neutrinos in the renormalisation
 group analysis of
masses, concentrating on its implications for lepton
mass matrices with a large $\mu - \tau$ mixing.
 We start by giving a  brief description of the procedure we are
following. We compute numerically the low energy values of
gauge and Yukawa couplings, taking  into account
two -- loop effects of the gauge couplings, one loop RGEs for
the Yukawas assuming an effective SUSY scale to account for low energy
threshold effects.

 First, we check the procedure by reproducing the standard results
 when no right handed neutrino is present in the theory.
 We start at the unification scale $M_U$ using as inputs $M_U$
 itself, the values  of the common gauge coupling $\alpha_U$, the top
 coupling $h_{t_G}$ and $h_{b_0}, h_{\tau_0}$.
  In obtaining the low energy values of $\alpha_{em}$, $a_3$,
 and $sin^2\theta_W$, we use the following ranges
 \begin{eqnarray}
 {\alpha_{em}}^{-1} = 127.9\pm .1 , \,\,
  a_3=.12\pm .01 ,\,\,
 sin^2\theta_W=.2319\pm  0.0004
 \end{eqnarray}
 The top mass $m_t$ is obtained in consistency with
 the correlation\cite{Lang}
 \begin{eqnarray}
 sin^2\theta_W(m_Z)=0.2324-10^{-7}\times
 \left\{ \left(m_t/GeV\right)^2-143^2   \right\}\pm 0.0003\label{eq:a}
 \end{eqnarray}
 We have converted this correlation into a relation
 between $sin^2\theta_W$ and $tan\beta$, using the
 relation of $m_{t}^{fxd}$ and $m_t$ , i.e.
 \begin{eqnarray}
 m_t = m_{t}^{fxd} sin\beta
 \end{eqnarray}
 We first search for the $tan\beta$ 's satisfying
 the above correlation. Then, this range is further constrained
 by the requirement $h_{b_0} =h_{\tau_0}$ at $M_U$.
 In the present work, we have concentrated  in the small $tan\beta$
 scenario, i.e. when $h_t \gg h_{b,\tau}$ and we
 comment for the large $tan\beta$ case later.

 At the next stage, we introduce the Dirac neutrino
 RGE and run all of them together from $M_U$ down to  the right
 handed neutrino scale $M_N$. We compare the predictions with
 those of the previous running (i.e. when there is no
 right--handed neutrino in the theory) and calculate the
 deviation from the equality of the $\tau - b$ unification for
 the same inputs at $M_U$.
 Below $m_N$ we add the RGE for the effective light neutrino
 Majorana mass matrix.
 We assume that we are in a diagonal basis, so we can run the three
eigenvalues of $M_N$ independently.

 Let us start with the low $tan\beta$ regime, assuming an effective
 SUSY scale $M_S^{eff}\le 1TeV$.
 We vary $a_U$ in a range close to a central value $\frac{1}{25}$
 and $M_U$ close to $10^{16}$ GeV.
 Our first observation is that the introduction of an intermediate scale
 where the right handed neutrino gets a mass, shifts  slightly the range
 of the parameter space for which  unification is possible.
 For example, assuming $h_{t_G}\approx 3$, i.e., close to its infrared fixed
 point, and assuming a unification point ranging in
  $M_G\sim (1.2-2.2)\times 10^{16}GeV$, with
 $\frac{1}{\alpha_U} \approx (23.81-25.64)$, the effective scale ranges from
 $M_S^{eff}=(100)GeV$ to $1 TeV$.
 Introducing the right--handed neutrino, we find $M_S^{eff}\ge 110GeV$.
  Some particular cases with the corresponding low energy predictions
are shown in Table 1.

\begin{center}
\vglue 0.2cm
\begin{tabular}{cccrccc}
\hline
   &    &      &     &        &             &    \\
$\frac{M_N}{GeV}$  &  $\frac{1}{\alpha_U}$ & $\frac{M_U}{10^{16}GeV}$ &
$M_S^{eff}$ & $\frac{1}{\alpha_{em}}$  &
$sin^2_{\theta_W}$ & $\alpha_3$ \\ \hline
   &    &      &     &        &             &    \\
$10^{16}$&23.81  &   2.18 &    100& 127.9&  0.2325&   0.121\\

&23.81  &   2.41& 110& 128.96&  0.2320&  0.123\\

&24.39&    1.97&  221& 128.09   &  0.2320&     0.120\\

&25.00&   1.46&  493& 127.98&   0.2321&     0.118\\

&25.64 &   1.08&    1212 & 127.82&    0.2319&     0.116\\
\hline
   &    &      &     &        &             &    \\
$10^{11}$& 23.81&  2.18&  110& 127.83& 0.2323&  0.122\\

& 23.81&  2.41  &  122& 127.90  &   0.2318&  0.124\\

& 24.39&  1.97&    270& 127.89&    0.2315&  0.122\\

& 25.00 & 1.46&    493& 128.05& 0.2321&   0.118\\

&25.64& 1.08&1212& 127.89&  0.2320&   0.115\\
 &  &      &     &        &             &    \\
\hline
\end{tabular}
\end{center}

 We should point out that, the presence of $\nu_R$ in the spectrum has
 little effect in the low energy values of $a_{em},\, sin^2\theta_W \, ,
 \alpha_3$ parameters.
 Moreover, for the above initial conditions
 the $sin^2\theta_W - m_t$ correlation, restricts $tan\beta$
 very close to unity  $tan\beta \le 2$.
 Of course, the biggest effects from the $\nu_R$ threshold are found
in the $b - \tau$ unification.

 For values in the perturbative regime of the top Yukawa coupling,
 $h_{t_G}$, at
$M_{GUT}$ we find it
impossible to obtain the correct $m_b, m_{\tau}$ masses starting
 with $h_b=h_{\tau}$ at $M_{GUT}$, even if
 the neutrino threshold is as high as $M_N = {\cal O}(10^{15}GeV$).
 For example, using  $h_{t_G}\approx 3.2$, (i.e. very close to its
 non-perturbative regime) and $h_{b_0}\approx h_{\tau_0}
\approx .0125$,  one can hardly
 obtain  a running mass for the bottom $m_b(m_b) \approx 4.5GeV$
 while the upper experimental bound is $m_b(m_b) \le 4.4GeV$.
 However, the solution of the solar puzzle needs $M_N \le 10^{13}GeV$.

 Therefore, in the following we do a complete two loop analysis and
calculate the exact deviation from the $b-\tau$ -- universality for
a reasonable range of the scale $M_N$. In our approach we first
 require the  $\tau$ -- mass  to be $1.749\pm 0.001$  at $m_Z$. Then we
search for the correct bottom mass and
 top mass as well as the required $tan\beta$.
 We choose the biggest possible coefficient
 for which we have a solution, which corresponds to a bottom
 mass $4.4 GeV$.
 The variation of this coefficient as a function of $M_N$
 is plotted in Fig {\it 1} ($h_{\tau_0} = 0.012$),
 for $h_{t_G} = 3.2$ and $2.0$ denoted in the plot
 by stars  and crosses respectively.
 For the rest of the input parameters we take:

\begin{eqnarray}
 M_S^{eff}=  544 GeV ,&
 M_{U} = 1.46 \times 10^{16} GeV ,&
 a_{U} = \frac{1}{25}
\end{eqnarray}
 We see, in agreement
 with the qualitative analysis,
 that for this parameter range and small
 $h_{t_G}$ it is not possible to obtain
 solutions for the $b - \tau$  ratio at unification
 being unity.
 The larger the Yukawa coupling for the top,
 the lower  the neutrino scale for which we find
 solutions.

It is useful to compare the mass and other parameter predictions
with respect to those obtained without the inclusion of $\nu_R$.
As has been pointed out in the previous section, in the presence
of $\nu_R$  correct predictions for $b-\tau$ masses
can be restored either by increasing $h_{\tau_0}$  or
by shifting $h_{b_0}$ to smaller values at $M_U$ with
a simultaneous change in the $\beta$ -- angle.
 In this latter case,  we
show in Fig {\it 2} the curves corresponding to the values of
$tan\beta$
 as a function of $M_N$, needed to compensate for the
decrease of the bottom mass. We find that as $M_N$
decreases there is a large effect
 on $tan\beta$, which drops for the two different
 choices for the top Yukawa coupling, from a common
 value of $1.35$ at $M_N \sim 10^{16}$GeV
 to $1.02$ and $1.13$
 for $h_{t_G} = 3.2$ and $2.0$ respectively,
 at $M_N \sim 10^{11}$GeV.
 This, combined with the running of the top Yukawa
 coupling to the fixed point (Fig 3),
 implies that we expect in this case a decrease
 in the top mass, as the qualitative description of the previous section
has indicated (Fig {\it 4}).
 The larger the initial value of the top Yukawa coupling
 and the smaller the initial value of $h_{\tau}$,
 the biggest the effect on $tan\beta$ through the running,
 and the larger the effect on the top mass.

 In Fig.
{\it 5} we see the effect of $M_N$ on $1/a_{em}$,
 which increases slightly as $M_{N}$ drops.
 At the same time, $sin^{2}\theta_{w}$ also
 increases slowly (Fig {\it 6}), while
 the strong coupling decreases (Fig {\it 7}).
 In all cases the effect is much smaller than the
 errors on these quantities,
 however it is enough to eliminate some of the
 solutions that were in the border
 of the  allowed region.

We would like to stress that,
in the case where the $h_{b_0}$
is the same while  $h_{\tau_0}$ is slightly increased
to compensate for the reduction caused by $\xi_N$,
there is no need to change the angle $\beta$.
{}For this reason  there is no significant effect
on the top mass, which preserves its value as
in the standard case.

{}Finally, in Fig {\it 8} we plot the light  $\tau$--Majorana
neutrino mass versus $h_{t_G}$ coupling, for three different values
of the heavy Majorana scale $M_N= 10^{12}$, $10^{13}$ and
$10^{14}$ GeV.

This analysis can be applied also in the case of the large $tan\beta$
regime. However in this case there are important corrections
to the bottom mass
 from one-loop graphs involving susy scalar masses and the $\mu$
 parameter. These graphs might
 induce  corrections to $m_b$ even of the order of $(30-50)\%$.

In view of the considerable uncertainties involved we have not extended the
numerical analysis to cover this case.

\section{Conclusions}

In this paper we have discussed the implications for low energy
physics of right-handed neutrinos  with masses at an
intermediate scale $M_N $.
For  $M_N\approx 10^{12-13}GeV$ (required
to give a $\tau$ neutrino mass ${\cal O}(1 eV)$ to serve as a hot
dark matter component) a $10\%$ deviation from
$b - \tau$ mass equality at the GUT scale
is needed to give an acceptable value for the
ratio $m_b/m_{\tau}$ as measured in the laboratory. We showed that it is
possible to
retain the $m_b^0=m_{\tau}^0$ GUT prediction of the $(3,3)$ --
elements of the corresponding mass matrices provided there is
sufficient mixing in the charged lepton mass matrix between
the two heavier generations.
The scenario we propose
can be realised in a simple
extension of the standard symmetry of electroweak interactions
to include a $U(1)$ family symmetry. Consideration of the
implications of this
symmetry for neutrino masses shows that the large mixing implied
 allows for a
simultaneous explanation of the atmospheric neutrino problem and the solar
neutrino problem.
This complements our previous discussion of
solutions to the solar
neutrino deficit while having a neutrino mass
in the range needed to fit the
COBE measurements in a hot plus cold dark matter universe. We have also
presented detailed numerical solutions of the renormalisation group equations
for the case of a heavy right-handed neutrino to support
the analytical analysis.





\newpage


\setlength{\unitlength}{0.240900pt}
\ifx\plotpoint\undefined\newsavebox{\plotpoint}\fi
\sbox{\plotpoint}{\rule[-0.200pt]{0.400pt}{0.400pt}}%


\vspace{0.2 cm}
\begin{center}
{\bf Fig. 8}
\end{center}
\vspace{1.0cm}
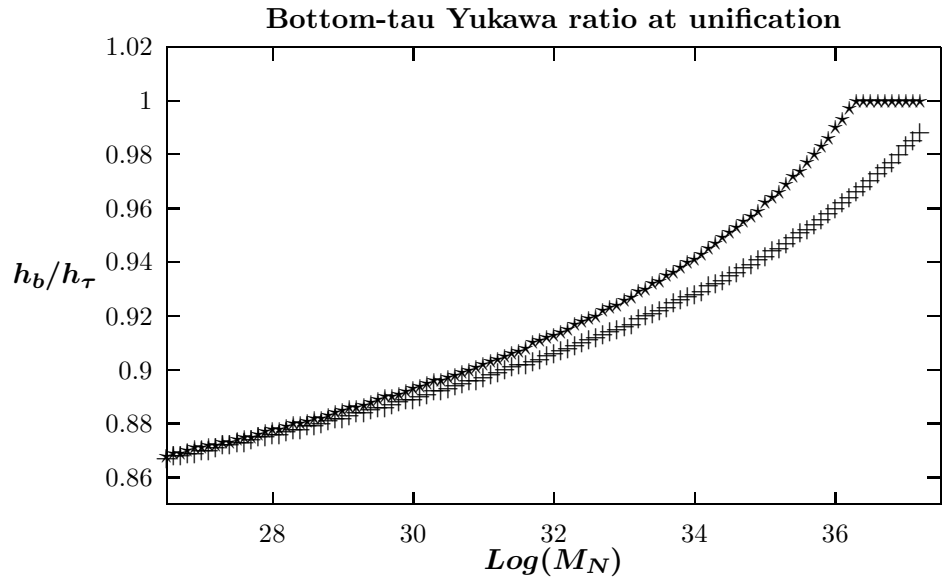
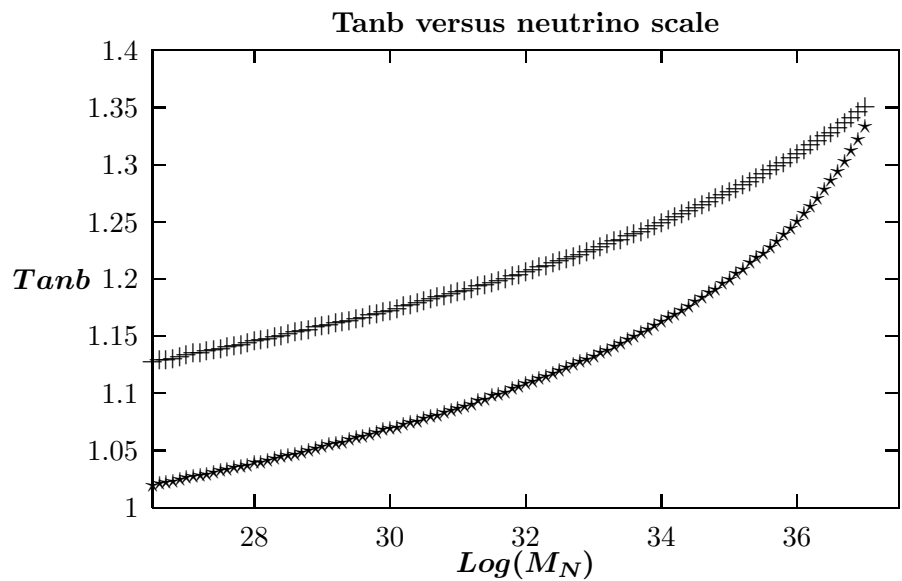
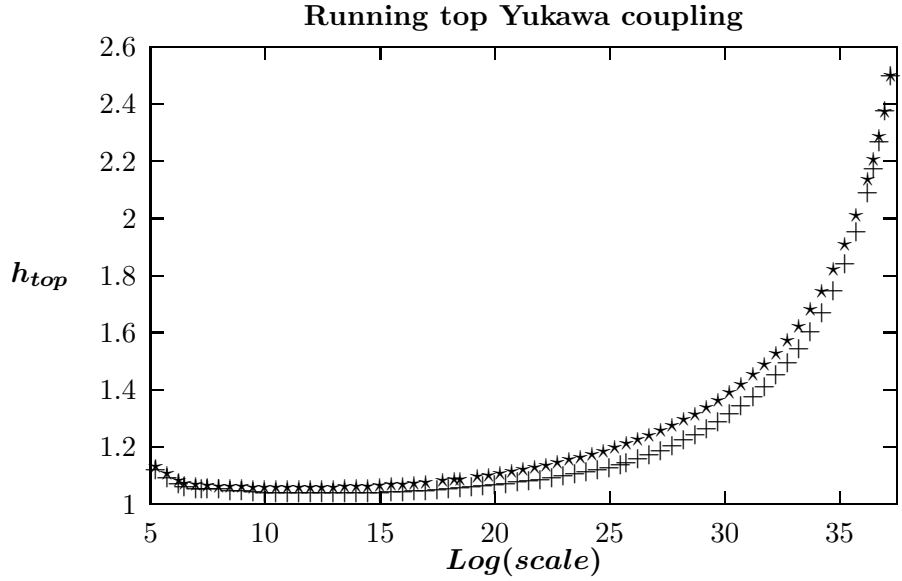
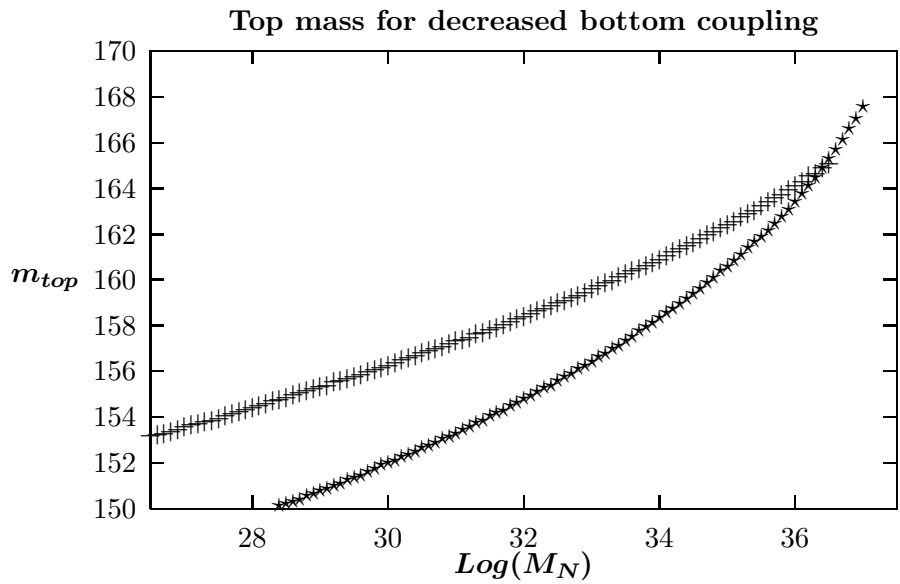
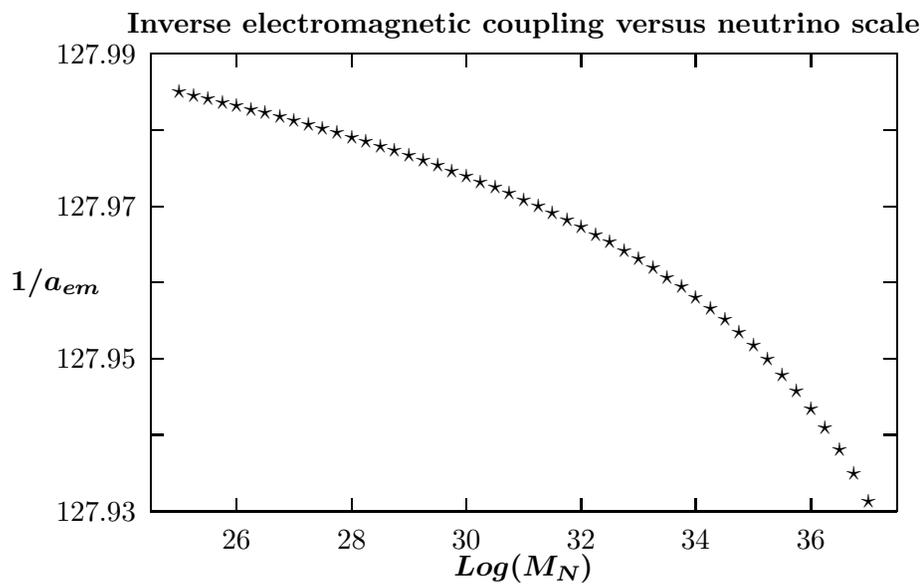
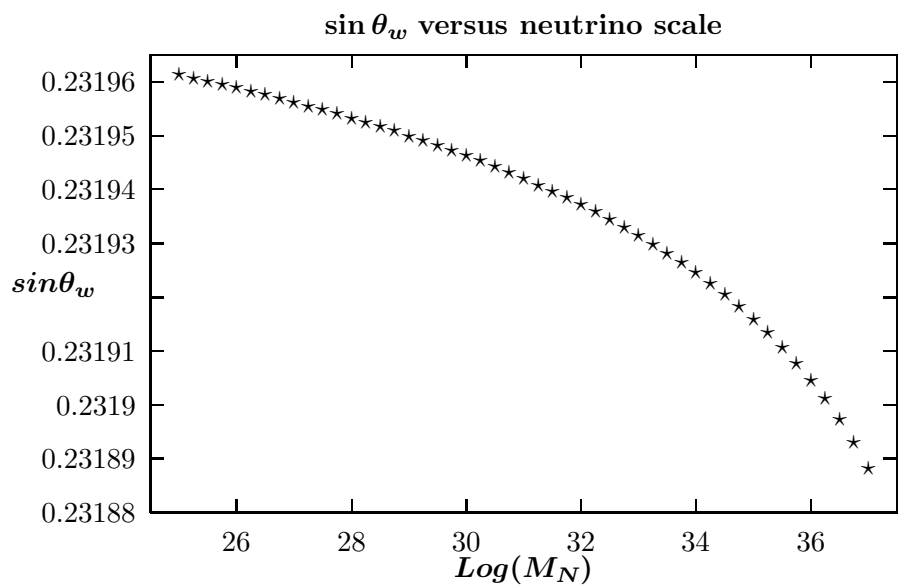
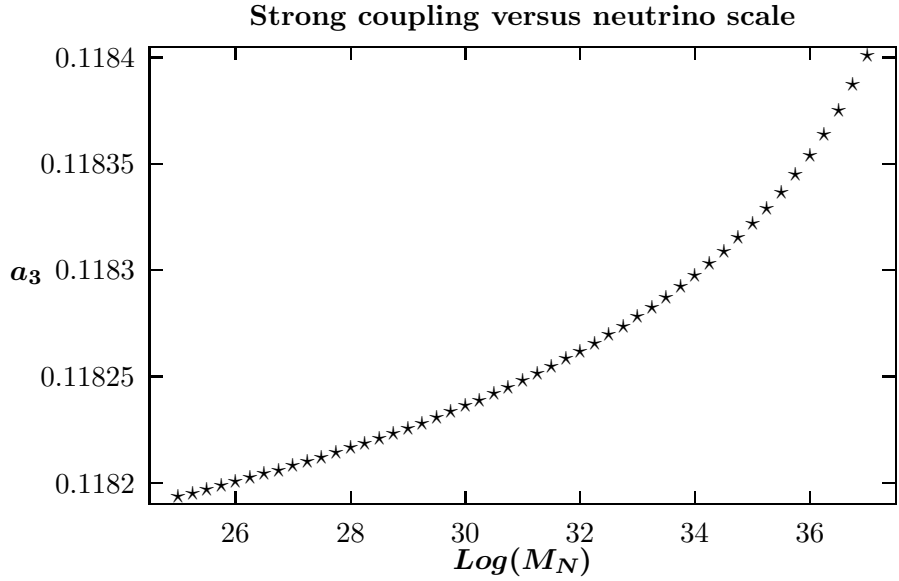
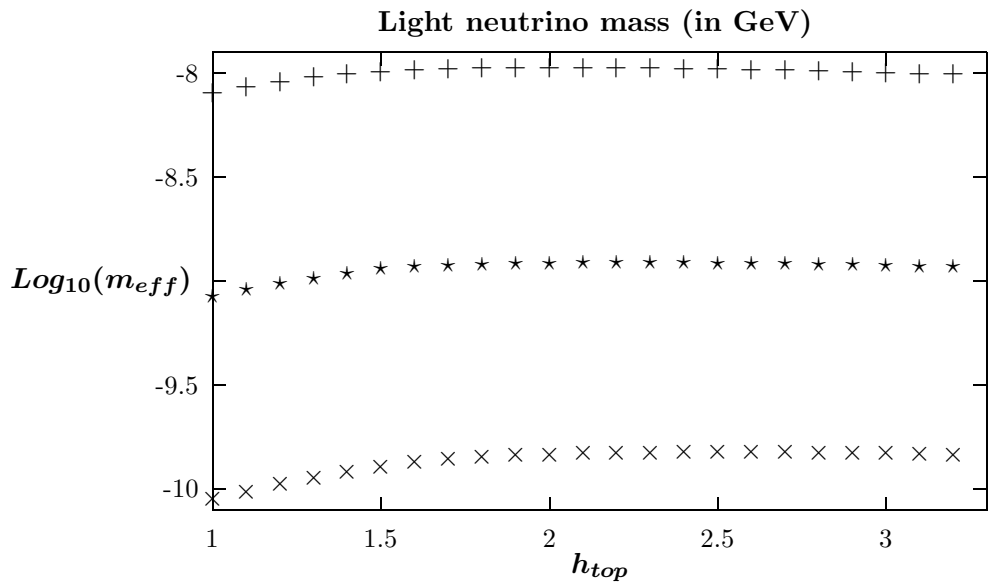
\end{document}